\documentclass[preprint,12 pt]{elsarticle}
\usepackage{natbib}
\usepackage{amssymb}
\usepackage{color}
\newcommand{\ped}[1]{\ensuremath{_{\rm #1}}}
\newcommand{\apex}[1]{\ensuremath{^{\rm #1}}}
\definecolor{blue}{rgb}{0,0,0}
\usepackage{float}
\usepackage[caption = false]{subfig}
\usepackage{graphicx}

\begin{document}

\title{Advanced surface characterization of Ba(Fe$_{0.92}$Co$_{0.08}$)$_2$As$_2$ epitaxial thin films}

\author [poli]{D. Daghero\corref{cor1}}
\ead{dario.daghero@polito.it}
\author [poli] {P. Pecchio}
\author [poli]{F. Laviano}
\author [poli]{R.S. Gonnelli}
\address[poli] {Dipartimento di Scienza Applicata e Tecnologia, Politecnico di Torino, Corso Duca degli Abruzzi 24, 10129 Torino, Italy}

\author[IFW]{F. Kurth}
\author[IFW]{V. Grinenko}
\author[IFW]{K. Iida}
\address[IFW]{Leibniz-Institut f\"{u}r Festk\"{o}rper-und Werkstoffforschung (IFW) Dresden, P.O.Box 270116, 01171 Dresden, Germany}

\begin{abstract}
We report on the systematic characterization of
Ba(Fe$_{0.92}$Co$_{0.08}$)$_2$As$_2$ epitaxial thin films on CaF$_2$
substrate in view of their possible use for superconducting electronic
applications. By using different and complementary techniques we studied
the morphological characteristics of the surface, the structural
properties, the magnetic response, 
and the superconducting properties in terms of critical temperature,
critical current, and energy gaps. Particular attention was paid to the
homogeneity of the films and to the comparison of their superconducting
properties with those of single crystals of the same compound.
\end{abstract}


\maketitle

\section{Introduction}
The study on Fe-based superconductors is one of the most fruitful fields of
present research in superconductivity. The rather high critical temperature
of some of these compounds, their high critical fields and critical currents,
the multiband electronic structure, the unprecedented sensitivity of the
superconducting properties to the structural parameters, are of great
interest for {\color{blue} both} fundamental reasons and possible applications in
transport and /or superconducting electronics \cite{johnston10}. Although the fundamental
research is traditionally focused on single crystals, the synthesis of
high-quality films is the indispensable step towards the fabrication of
superconducting electronic devices that, in most cases, have Josephson
junctions as their basic constituting element \cite{kadin99,seidel11}. Moreover, thin films are more suitable than single crystals for investigation of transport and optical
properties, thanks to their dimensionality and the possibility to control
their shape by lithography. Additionally, thin films are also important for
fundamental research, since they allow exploring the effects of lattice
deformation by strain/stress (hardly accessible in single crystals) on
structural parameters, surface morphology, transport and fundamental
properties.

The growth of epitaxial thin films of Co-doped BaFe$_2$As$_2$  on different
substrates by using a simple pulsed-laser deposition (PLD) technique has been
demonstrated since 2009 \cite{katase09, iida09, lee10b, engelmann13}. A strong effect of the substrate on the critical temperature was observed, due to strain-induced modification of
the ratio $c/a$ of the cell parameters. Recently, the use of
single-crystalline CaF$_2$ as a substrate has been shown to allow the growth
of epitaxial films of Ba(Fe$_{1-x}$Co$_x$)$_2$As$_2$ with a record high
critical temperature of nearly 28 K \cite{kurth13b}.

In this paper we present a systematic characterization of these  films by
means of various complementary  techniques in order to test (and assess)
their quality from the points of view of structure (epitaxy, texture),
surface morphology (granularity, roughness), and superconducting properties via magnetic,  transport and spectroscopic measurements. In particular, we compare the results of
local and non-local techniques in order to check the homogeneity of the
films; moreover, we compare the superconducting properties (critical
temperature and gap amplitudes) to those of single crystals. The results
provide thorough information on the features of these films, which is a basic
step prior to their use for advanced fundamental studies and applications.

\section{Thin film growth}
The Ba(Fe$_{1-x}$Co$_{x}$)$_2$As$_2$ thin films, of thickness  $t= 50$ nm,
were deposited on (001) CaF$_2$ substrates by PLD \cite{kurth13b} in ultra high vacuum (base pressure of $10^{-9}$ mbar) using a Ba(Fe$_{1-x}$Co$_{x}$)$_2$As$_2$ polycrystalline target with cobalt content $x=0.08$. The high phase purity of the target was verified by means of powder
X-ray diffraction (XRD) in Bragg-Brentano geometry \cite{kurth13b,kurth13a};
before the film deposition the target surface was cleaned by about 1000 laser
pulses. The CaF$_2$ substrate was cleaned in an ultrasonic bath by using
acetone and isopropanol, and then subjected to an additional heat treatment
at 700$^{\circ}$C for 30 minutes. Because of the strain induced by the substrate, it is
impossible to {\color{blue} determine} the Co content of the films by measuring the lattice parameters and comparing the results with single crystal data, as instead is commonly done in the case of the bulk material. The Co content was therefore measured directly by means of energy dispersive X-Ray (EDX)
spectroscopy (with an EDX Oxford X-ACT spectrometer). Different spectra were
taken on the same film, either over areas of a few {\color{blue} square microns}, or in single
points. Because of the very small thickness of the films as compared to the
depth probed by EDX, a large signal from the CaF$_2$ substrate was always
observed. Nevertheless, the compositions determined from all these spectra
turned out to be well consistent with one another. On averaging over all the
spectra, the value  $x = 0.074\, \pm \, 0.017$ was obtained for the Co
content, in very good agreement with the nominal value $x\ped{nom}=0.08$. The uncertainty
$\delta x=0.017$ is here equal to one half of the spread of $x$ values
obtained in single spectra.

\section{Surface morphology}
\begin{figure}[h]
\begin{center}
\includegraphics[keepaspectratio, width=0.8\textwidth]{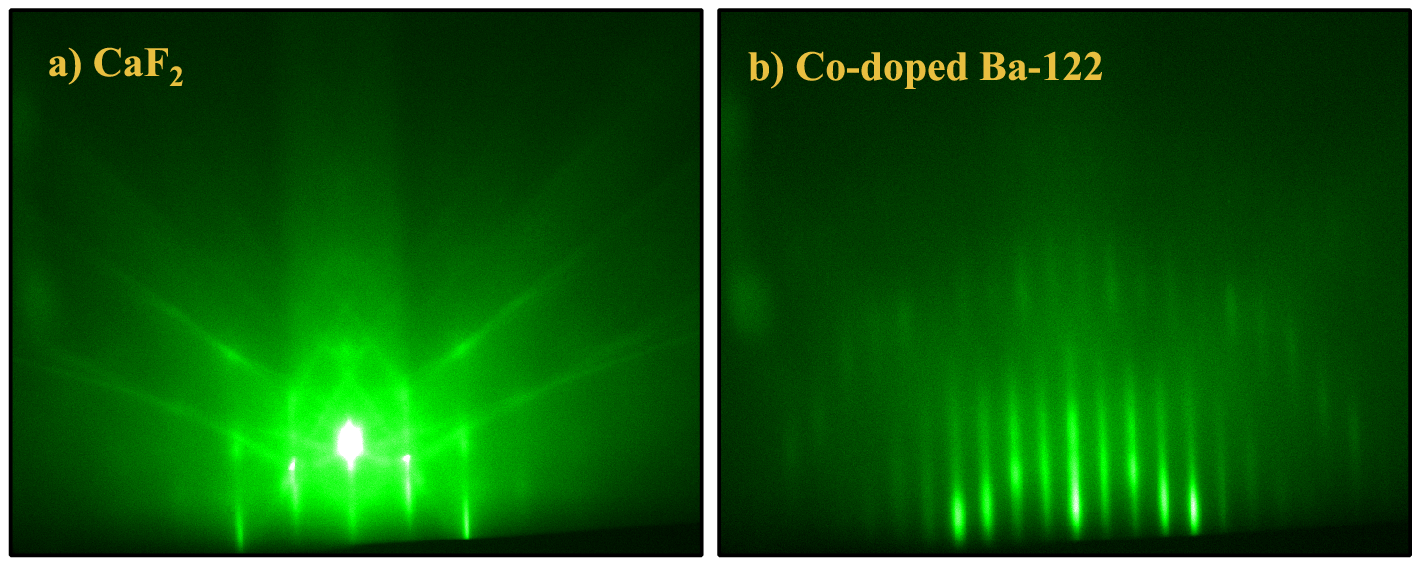}
\end{center}
\caption {} \label{fig:RHEED}
\end{figure}

Figure \ref{fig:RHEED}a shows a reflection high-energy electron diffraction
(RHEED) image of the CaF$_2$ substrate after cleaning and thermal treatment.
Here the incident electron beam is parallel to the [110] direction of the
CaF$_2$ substrate. It is clear from this figure that streaks and well distinct Kikuchi lines are observed, which indicate a smooth and single-crystal surface of CaF$_2$ substrate. A RHEED image of the Co-doped Ba-122 thin film grown on this substrate, taken just after
the deposition, is shown instead in Figure \ref{fig:RHEED}b. The pattern in this
case shows long streaks perpendicular to the shadow edge, and centered at
positions on the Laue circles. This is typical for a multilevel surface, i.e.
for a high number of grains with smooth terraces.

The morphology suggested by the RHEED is confirmed  by direct imaging of the
film surface by field-emission scanning electron microscopy (4248 Merlin
ZEISS FESEM). Figure \ref{fig:FESEM}a shows a secondary-electron (SE) image
taken over an area of about $1.8 \times 1.8 \, \mu \mathrm{m}^2$. This image
mainly carries morphological / topographical information; flat and smooth
terraces, with parallel edges of about $100 - 150$ nm in size are clearly
visible, which protrude from a pattern of well connected grains. This further
indicates that Co-doped Ba-122 films grow in a terraced island mode. Figure
\ref{fig:FESEM}b reports a greater-magnification (i.e. smaller-scale) image
taken on a different region of the film surface, and Figure \ref{fig:FESEM}c
shows a backscattered-electron (BSE) image of the same area. A diagonal
stripe-like pattern not related to the granular structure is very clearly
seen in this picture (and also, though less clearly, in panel (b)); this
pattern might be associated with correlated defects along the $c$-axis as
also discussed in section \ref{sec:transport}.

Additional and complementary information about the surface morphology was
collected by means of atomic-force microscopy (AFM) measurements. Figure
\ref{fig:AFM}a and b show two AFM images taken on the same region of the
film, though on a different scale. The correspondence between color and
height is shown by the vertical bar on the right side of each picture. The
flat terraces with parallel edges already observed by FESEM are very clearly
visible here; the additional information is that their top surface is flat
but not perfectly horizontal. The distribution of heights calculated over the
whole area shows a broad maximum at 5 nm, which indicates that the terraces
protrude by (on average)  5 nm out of the underlying surface. The root mean square (RMS)
roughness is 1.15 nm while the size of the terraces, determined by means of
$z$-height profile cuts, ranges between 50 and 200 nm.  Figure \ref{fig:AFM}c
shows the same image as in (b) but after a Prewitt gradient filtering, that
allows highlighting the elevation of the terraces {\color{blue} above} the background.

\section{Structural properties}\label{sec:structure}
A structural characterization of the films was carried  out by means of X-ray
diffraction spectroscopy. Figure \ref{fig:XRD}a shows a $\theta-2\theta$-scan
in the Bragg-Brentano geometry, that demonstrates that the films grow free
from impurities and secondary phases. Moreover,  only the $00\ell$
reflections from Co-doped Ba-122 and of the CaF$_2$ substrate are recorded,
which proves that the film is $c$-axis oriented. Figure \ref{fig:XRD}b
reports  $\phi$-scans of the 112 reflection of Co-doped Ba-122 film (top
curve). Strong and sharp peaks (average $\Delta\phi$ value of $0.73^{\circ}$)
are observed every $90^{\circ}$, indicative of four-fold symmetry. The bottom
curve in the same figure is a $\phi$-scan of the 111 CaF$_2$ reflection,
reported here for comparison and to confirm the epitaxial relation between
the film and the substrate.  Note that the CaF$_2$ substrate peaks appear
$45^{\circ}$ away from the Co-doped Ba-122 peaks. This clearly indicates
that the Co-doped Ba-122 film is biaxially textured and its epitaxial
relation with the substrate is (001)[110] Co-doped Ba-122 $\parallel$
(001)[100] CaF$_2$.

\section{Transport and magnetic properties} \label{sec:transport}
Standard four-probe resistance measurements were performed in a $^4$He
cryostat to determine the transport critical temperature and the transition
width. The measurements were conducted by using either the standard collinear
configuration (with contacts along the diagonal of the 5 mm $\times$ 5 mm
film) or the van der Pauw (vdP) configuration, with four contacts on the
corners (see the left inset to Figure \ref{fig:RESISTANCE}a). With respect to the
collinear configuration, the vdP one allows probing a much larger region of
the sample surface, and thus provides a well-representative measure of its
average normal-state properties. In other words, if the measured resistivity is written as
a weighted average over the entire sample surface, i.e. $\langle \rho
\rangle=\int \rho(x,y) g(x,y) dx dy $ where $g(x,y)$ is a position-dependent
weighting function, the vdP configuration allows maximizing the geometrical
extension of the region (the so-called ``sweet spot'' \cite{koon96,koon98})
where the weighting function is maximum. In the case of a square sample with
the contacts on its corners, as in our case, the van der Pauw resistance
measurements probe an extended region of the film (of the order of some {\color{blue}square millimeters}) \cite{koon98}.

A typical result of a $R(T)$ measurement up to 250 K is shown in the right
inset to Figure \ref{fig:RESISTANCE}a (solid line). Here the current was flowing
through the two top contacts and the voltage drop was measured across the
bottom ones (see left inset). None of the curves measured in
the films showed anomalies that could suggest the existence of macroscopic
inhomogeneities either in the transition temperature or in the normal-state
conductivity \cite{vaglio93,mosqueira94}. In the normal state, the $R(T)$
curve displays a minimum around 75 K. A comparison with the resistivity curves
of single crystals \cite{chu09} shows that this shape is typical of the
underdoped region, while in optimally-doped ($x=0.061$) and overdoped
crystals the $R(T)$ curve is monotonic. To clarify this point, the right inset
to Figure \ref{fig:RESISTANCE}a shows the resistance of the film compared to the resistivity of a
single {\color{blue} crystal} with $x=0.051$ (dashed line), both normalized to their values at 250 K. Further indication that our films lie in the slightly underdoped region of the phase diagram comes from the fact that in the films deposited on CaF$_2$ substrate the highest $T\ped{c}$ is achieved for $x=0.10$ \cite{pecchio13}. It is clear therefore that the phase diagram of these films shows a superconducting dome which is shifted to higher doping (and also extends to higher temperature)
with respect to that of single crystals. This fact has already been mentioned
in our previous work \cite{pecchio13}. The confirmation of the Co content
of the films provided by EDX (see above) rules out a systematic
overestimation of the doping, so that the only factor responsible for the
wider superconducting dome of films is the effect of the substrate
\cite{kurth13b}.

Let us focus now on the superconducting transition. The resistance drops to
90\% of its residual value at $T\ped{c}^{90}=25.2$ K and to 10\% at
$T\ped{c}^{10}=23.5$ K (see main panel of Figure \ref{fig:RESISTANCE}a).
The zero-resistance state is reached at $T\ped{c}^0= 22.9$ K.
Resistance measurements over a whole set of films with the same doping
content (but from different batches) {\color{blue}have} given the average values
$T\ped{c}^{10}=(23.85 \pm 0.35)$ K and $T\ped{c}^{90}=(25.4 \pm 0.2)$ K.

As mentioned above, the resistance measurement is definitely not a local one; and also the
critical temperature is averaged over a large portion of the film. To check
whether this measurement is representative of the \emph{local} properties, we
used a technique which is very seldom employed for that purpose, i.e.
point-contact Andreev-reflection spectroscopy (PCARS) \cite{daghero10}. This
consists in measuring the differential conductance of a point contact between
a normal metal and the superconductor under study, as a function of the
voltage applied across the contact itself. If the contact is small enough
(i.e. smaller than the electronic mean free path) the $dI/dV$ vs $V$ curve
shows a conductance enhancement due to Andreev reflection \cite{andreev64},
that contains information about the amplitude and symmetry of the
superconducting gap. Since the latter closes at the \emph{local} critical
temperature of the region where it is measured, PCARS can provide point-like
measurements of the critical temperature, defined here as the temperature
$T\ped{c}\apex{A}$ at which the features associated to Andreev reflection disappear and
the normal-state conductance is recovered.

To make the point contacts on the films \cite{pecchio13}, we used a thin Au
wire ($\varnothing=18 \,\mu\mathrm{m}$) kept in electrical and mechanical
contact with the film surface in one single point, thanks to a small drop of
Ag conducting paste. Although the diameter of the drop is rather large
($\varnothing \leq 100 \,\mu\mathrm{m}$), the effective contact occurs only
here and there between the film surface and single Ag grains, naturally
selecting the more conducting channels within a microscopic region. The
spectrum of each point contact is thus actually an average (over a
micrometric scale) of the signal of several nanometric contacts.

A comparison between the values of $T\ped{c}\apex{A}$ we measured in different regions
of a given film (hollow symbols) with the relevant $R$ vs $T$ curve (red filled symbols) is shown in the main panel of Figure \ref{fig:RESISTANCE}. Here
the abscissa of each hollow circle is the local $T\ped{c}\apex{A}$ of a point contact;
since each PCARS spectrum must be acquired at thermal equilibrium, the
uncertainty on $T\ped{c}\apex{A}$ comes from the temperature step between the first
normal-state spectrum and the last superconducting one. The ordinate of the
points does not have a special meaning (it has been adjusted so that the
points are superimposed to the $R(T)$ curve for ease of comparison). It is
worth noting that all the values of the local $T\ped{c}\apex{A}$ are close to
$T\ped{c}\apex{90}$ or lie between this temperature and $T\ped{c}\apex{10}$.
The absence of $T\ped{c}\apex{A}$ values that fall in the lower $10\%$ of the resistive
transition is an experimental evidence common to all the measured films.
This fact certainly indicates that there is no significant heating in the contact
region (i.e. the temperature of the contact is not higher than that of the
bath), as instead would happen if the contacts were not in the proper
spectroscopic regime.

The magnetic properties of the films were investigated in a Quantum Design SQUID magnetometer. The temperature dependence of the normalized magnetic moment $m(T)/m(1.8 \mathrm{K})$ is shown in Figure \ref{fig:RESISTANCE}b (blue {\color{blue}open} symbols). In the experiment the film was cooled in zero magnetic field down to $T$ = 1.8 K, then a magnetic field $H=$ 5 Oe parallel to the $ab$ plane of the film was applied, and the superconductor was heated up above $T\ped{c}$ giving the zero-field-cooling (ZFC) curve in Figure \ref{fig:RESISTANCE}b. After this, the film was again cooled down to $T$ = 1.8 K in the presence of $H\parallel ab =$ 5 Oe, and the field-cooling (FC) curve in Figure \ref{fig:RESISTANCE}b was measured.
The splitting between ZFC and FC curves corresponds to the temperature $T\ped{c}^{0}$ at which non-zero critical current appears in the film volume. In the case of the homogeneous superconducting state $T\ped{c}^{0}$ exactly corresponds to zero resistance. Usually, from the value of the diamagnetic signal at low temperatures (ZFC curve) it is possible to estimate the superconducting volume fraction in the sample. In our case the film thickness $t$ is smaller than the magnetic penetration depth $\lambda$ \cite{johnston10}, therefore the value of ZFC signal does not correspond to the actual superconducting volume. However, the pronounced diamagnetic signal and the high transport critical currents (see below) indicate that a large volume fraction of the film is in the superconducting state at $T < T\ped{c}^{0} \approx$ 23 K.

To study the transport properties of the films in greater detail, we also measured the in-field critical current density $J\ped{c}$ as a function of temperature and magnetic field. The measurements were carried out by using a standard 4-probe collinear configuration. A voltage criterion of 1\,$\rm\mu V/cm$ was employed for evaluating $J\ped{c}$. The magnetic field $H$ was applied in maximum Lorentz force configuration during all measurements. In Figure \ref{fig:CURRENTDENSITY}a, the field dependence of $J\ped{c}$ is shown for
both main crystallographic orientations of the magnetic field (i.e. $H
\parallel c$ and $H \parallel ab$)  in a temperature range between 10 and 22 K.
In the low-field regime, $J\ped{c}$ is very little dependent on the field
direction (i.e. the critical current density is isotropic). This is even
clearer by looking at the angular dependence of $J\ped{c}$ shown in Figure
\ref{fig:CURRENTDENSITY}b. Here the angular dependence of the critical
current is very weak (i.e. $J\ped{c}$ is almost isotropic) up to magnetic
field intensities of the order of 5 T. Actually $J\ped{c}$ for $H \parallel
c$ is slightly larger than that for $H \parallel ab$ below 3 T, which is
unexpected from the electronic anisotropy of this system \cite{iida10b,putti10}. Such an inverse correlation of $J\ped{c}$ anisotropy strongly indicates the presence of defects along the $c$ axis, that may be related to the irregular stripe-like pattern observed by FESEM in BSE
mode.

\section{Local energy gaps}
The determination of the local gap values is not a standard surface characterization
technique for films; however, the homogeneity of the gap amplitude over the
surface is a useful indication if the films are meant to be used for
electronic applications (especially in the long-term purpose of realizing
circuits with multiple Josephson junctions, large-area SQUIDs and so on).

The values of the gaps were determined by means of PCARS in the same point
contacts used for the measurement of the local $T_c^A$ (see left inset to Figure
\ref{fig:RESISTANCE}a). Figure \ref{fig:PCARS}a shows two representative
examples of normalized PCARS spectra measured at 4.2 K. In general, the
normalization is obtained by dividing the raw spectrum ($dI(V)/dV$ vs $V$) of
the normal-metal/superconductor (NS) junction by the same spectrum obtained
when both the banks are in the normal state. Because of well-known effects
associated to the large resistance of the films \cite{chen10b}, however, the
normal-state spectrum in our case is shifted downwards (see for example
\cite{pecchio13} and \cite{doring13}) and cannot be used for this purpose.
The normalization was therefore done with respect to the polynomial fit of
the high-energy tails of the spectrum. This procedure gives rise to some
uncertainty on the amplitude of the signal but does not affect the position of the structures associated to the energy gaps.  In order to extract the gap values, the normalized spectra (open symbols) were fitted by using the standard two-band 2D BTK model \cite{BTK,kashiwaya00} with two isotropic gaps {\color{blue} whose amplitudes will be called $\Delta\ped{L}$ (the larger) and $\Delta\ped{S}$ (the smaller)}.
This is consistent with ARPES results in Co-doped Ba-122
\cite{terashima09} but also with the results of PCARS in single crystals of
the same compound \cite{tortello10} at optimal doping. The result of the fit
is shown as solid lines in Figure \ref{fig:PCARS}a. Since the model contains
7 parameters, the result is not univocal, i.e. there is a range of parameters
that allows obtaining quantitatively acceptable fits within a certain
interval of confidence. The gap values indicated in the labels are actually
the midpoints of the range of acceptable {\color{blue} $\Delta\ped{S}$ and $\Delta\ped{L}$ } values, and the uncertainty expresses the half-width of that range. Figure
\ref{fig:PCARS}b reports the values of the gaps obtained in different point
contacts. The values are rather well compatible with one another; in
particular, a good homogeneity of the small gap is observed, while the large
gap shows greater spread: it varies within a range of about 3 meV including
the error bars (this spread is however not unusual in Fe-based compounds \cite{daghero10}). Note that, as shown elsewhere \cite{pecchio13}, the gap amplitudes shows no dependence on the contact resistance, that here varies by over one order of magnitude.
The values of the gaps obtained in these films may be compared to those obtained in single crystals. In general, this comparison can tell whether the superconducting properties of the bulk are weakened in thin films. Here, because of the strong effect of the substrate, thin films are structurally different from the bulk material so that this comparison must be taken more cautiously. Anyway, just for the sake of discussion, the gap values measured in $c$-axis contacts in single crystals with critical temperature (measured by transport) $T\ped{c}\apex{onset}=24.5$ K turned out to be $\Delta\ped{S}=4.1 \pm 0.4$ meV and $\Delta\ped{L}=9.2 \pm 1.0$ meV  \cite{tortello10}. These ranges are indicated in Figure \ref{fig:PCARS}b by dashed bands. It is clear that the values of $\Delta\ped{S}$ and $\Delta\ped{L}$ obtained in crystals lie well within the range of measured amplitudes in films, although in the latter case the spread is larger for both the small and the large gap. It is also true, however, that single crystals were much smaller than the films (the top surface area was at most $1\times 1 \, \mathrm{mm}^2$) and that an extensive study of the spatial homogeneity of the gaps in these samples was not performed.

\section{Conclusions}
Co-doped Ba-122 thin films of thickness $t=50$ nm were prepared by pulsed laser deposition on (001)CaF$_2$ single crystalline substrates and morphologically, structurally and electro-magnetically characterized by means of various experimental techniques. RHEED, FESEM and AFM analyses show that the films grow in a terraced island mode, with flat rectangular terraces of about 100 nm in size protruding by about 5 nm out of a background of well-connected grains. The surface is very smooth and homogeneous, with a small RMS roughness of 1.15 nm.
As demonstrated by $\theta-2\theta$ XRD scans, the films are free from impurities and secondary phases, and are $c$-axis oriented; $\phi$ scans also show that they are highly biaxial textured, with the Ba(Fe$_{1-x}$Co$_x$)$_2$As$_2$ cell rotated by 45$^{\circ}$ with respect to that of the CaF$_2$ substrate. The actual cobalt content, measured by EDX, is $x=0.074 \pm 0.017$, in good agreement with the nominal one $x\ped{nom}=0.08$. Nevertheless, these films are slightly underdoped (indeed, the maximum $T\ped{c}$ is achieved for $x=0.10$) as also witnessed by the minimum in their $R(T)$ curve. The discrepancy with the case of single crystals (where optimal doping occurs at $x=0.061$ \cite{chu09}) is thus likely to arise from the strong compressive strain (and the consequent enhancement of the $c/a$ ratio) induced by the CaF$_2$ substrate on Co-doped Ba-122 \cite{kurth13b}.
The critical temperature determined by electric transport measurements is indeed by almost 2 K higher than that of single crystals with the same cobalt content. The local surface critical temperature $T\ped{c}\apex{A}$ determined by point-contact spectroscopy in different regions of a given film turns out to be in perfect agreement with the resistive transition; moreover, evidence for bulk superconductivity is given by magnetization measurements, which also show the correspondence between the onset of a diamagnetic signal and the achievement of the zero-resistance state. The superconducting gaps (in particular the smaller one, {\color{blue} $\Delta\ped{S}$,} which generally plays a major role in tunneling and presumably also in Josephson effect \cite{seidel11}) are rather homogeneous over the surface on a millimetric scale, and their amplitudes are well compatible with those measured in single crystals of the same compound. All these features are fundamental in view of applications in superconducting electronics. Finally, the rather high critical current density  and its small anisotropy (promising for magneto-transport applications) suggest the presence of defects along the $c$ axis, acting as pinning centers for $H \parallel c$, as also shown by FESEM BSE imaging.
In conclusion, our Co-doped Ba-122 thin films show robust and homogeneous superconducting properties -- in some respect analogous to those of single crystals -- which are very promising for fundamental studies and superconducting applications, although some focused optimization of these properties can be envisaged in view of their use for specific purposes.

\section{Acknowledgements}
The authors wish to thank Mauro Raimondo for FESEM and EDX measurements.
This work was done under the Collaborative EU-Japan Project ``IRON SEA'' (NMP3-SL-2011-283141).

\newpage

\section{Figure Captions}

\textbf{Figure 1}  \\ 
RHEED images of (a) the CaF$_2$ substrate and (b) the 8 \% Co-doped Ba-122 thin film. In the latter case, the picture was acquired after the deposition at room temperature. As discussed in the text, the basal plane of Co-doped Ba-122 is rotated by $45^{\circ}$ with respect to the substrate. Hence the electron beam is here parallel to the [100] direction of the Co-doped Ba-122.

\bigskip
\textbf{Figure 2} \\ 
FESEM images of the surface of a 8\% Co-doped Ba-122
thin film deposited on CaF$_2$ substrate. (a,b) Two representative
secondary-electron (SE)  images with different magnifications
(the scale is indicated in the labels). Clear, regular, flat terraces of
rectangular shape (side length approximately 100 nm) are seen that
protrude from an array of well-connected grains. (c) Backscattered-electron
(BSE) image of the same area as in (b), that shows a diagonal stripe-like
pattern not related to the granular structure, and possibly indicating the
existence of correlated defects parallel to the $c$ axis (i.e. perpendicular
to the plane of the figure).

\begin{figure}[t]
\begin{center}
\includegraphics[keepaspectratio, width=\textwidth]{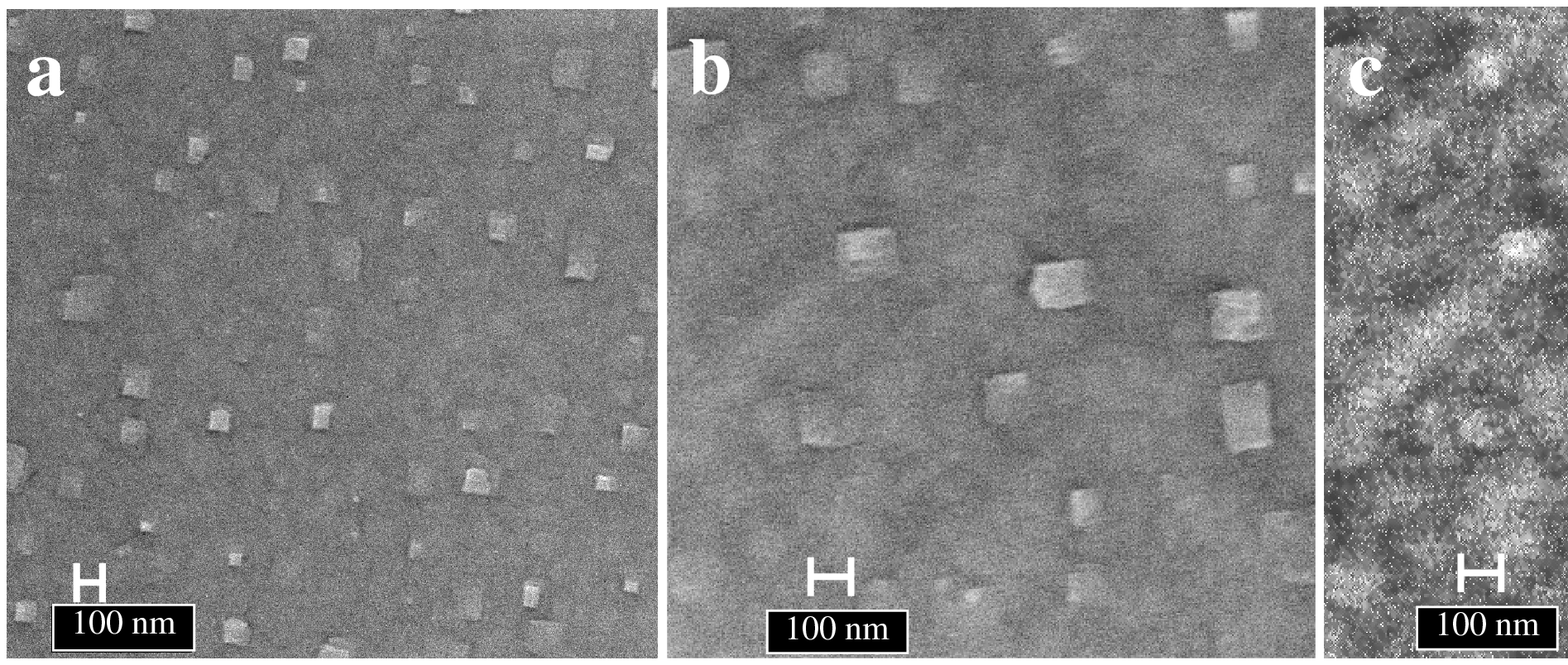}
\end{center}
\caption {} \label{fig:FESEM}
\end{figure}

\begin{figure}[t]
\begin{center}
\includegraphics[keepaspectratio, width=\textwidth]{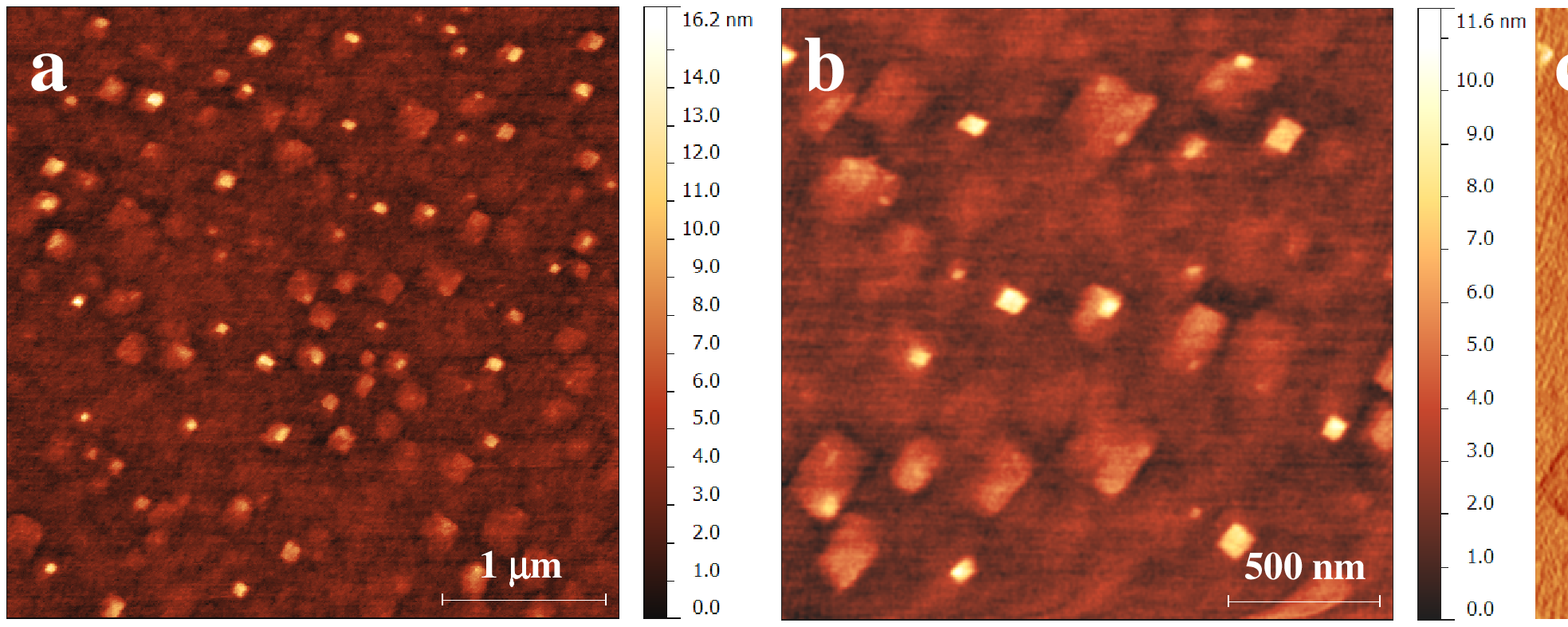}
\end{center}
\caption{} \label{fig:AFM}
\end{figure}

\begin{figure}[t]
\begin{center}
\includegraphics[keepaspectratio, width=0.8\textwidth]{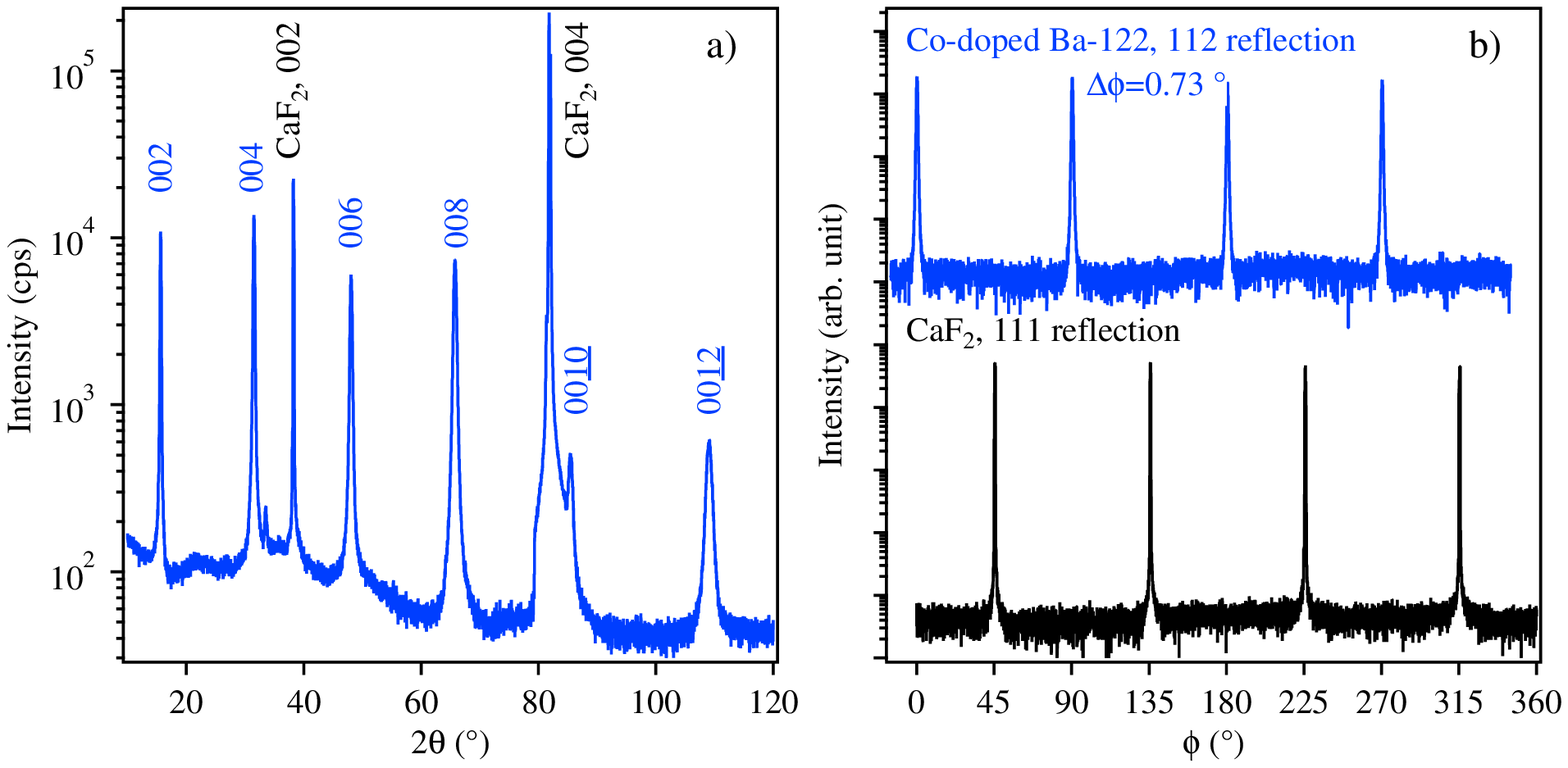}
\end{center}
\caption{}
\label{fig:XRD}
\end{figure}

\begin{figure}[t]
\begin{center}
\includegraphics[keepaspectratio, width=0.8\textwidth]{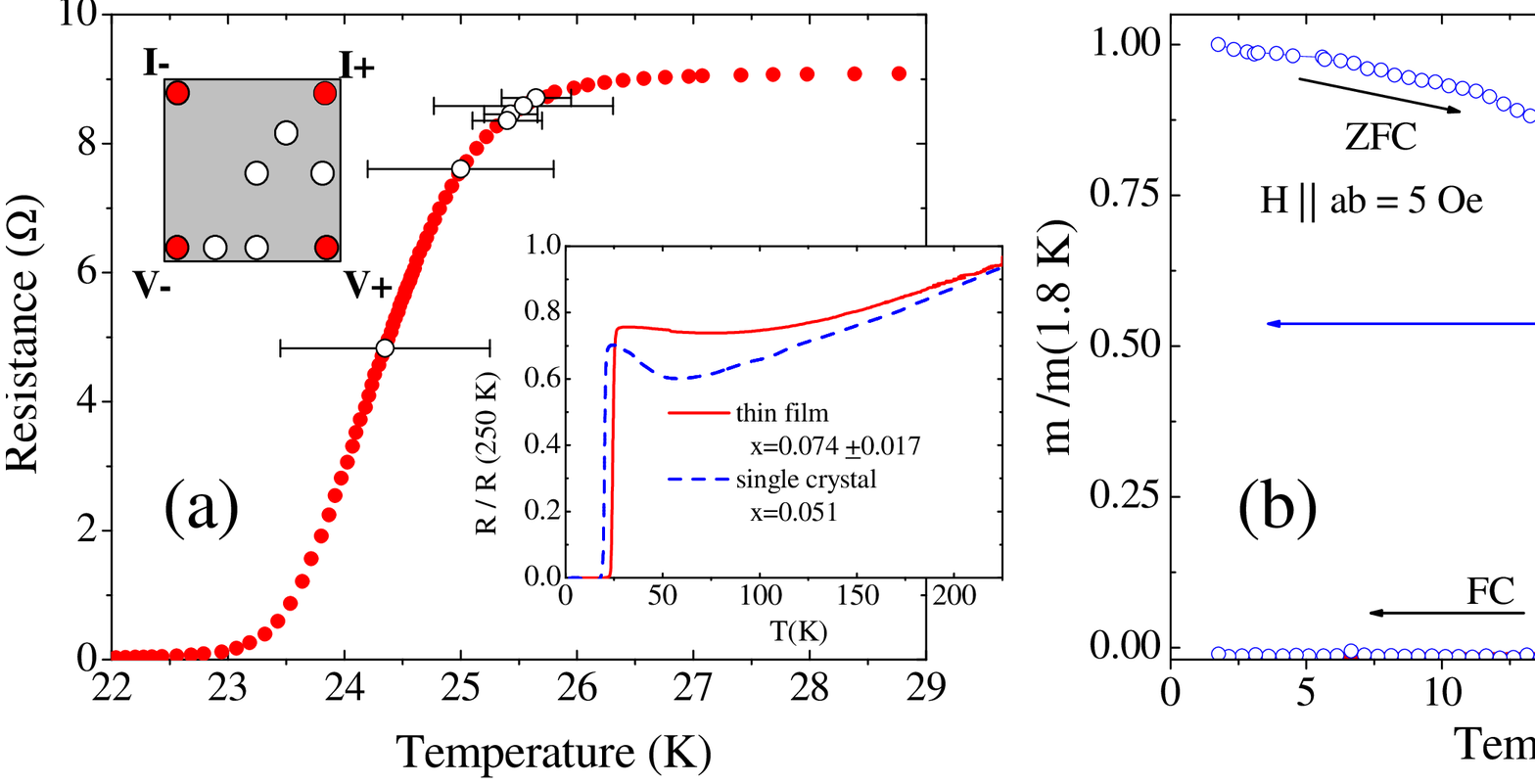}
\end{center}
\caption {} \label{fig:RESISTANCE}
\end{figure}

\begin{figure}[ht]
\begin{center}
\includegraphics[keepaspectratio, width=0.8\textwidth]{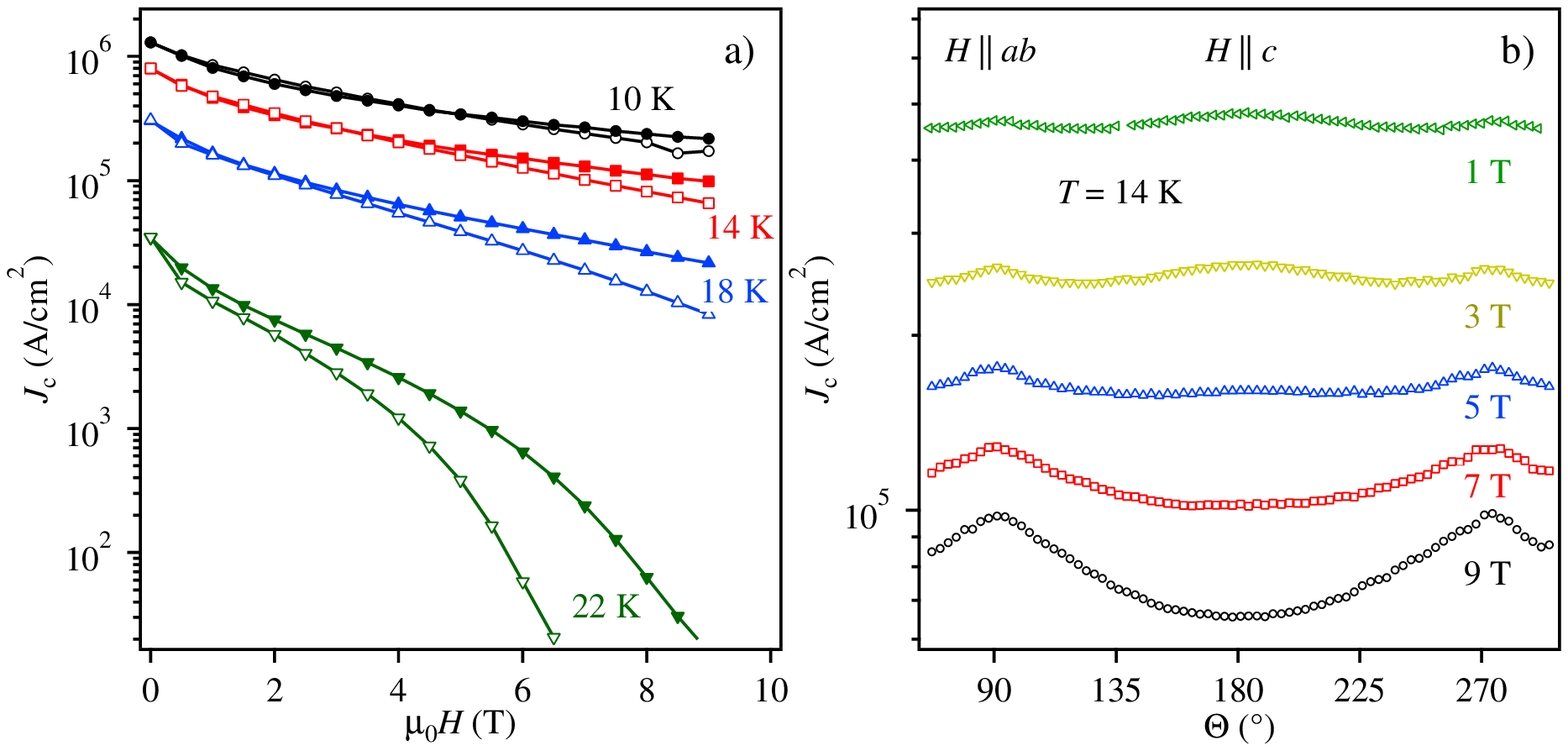}
\end{center}
\caption{}
\label{fig:CURRENTDENSITY}
\end{figure}

\begin{figure}[ht]
\begin{center}
\includegraphics[keepaspectratio, width=0.6\textwidth]{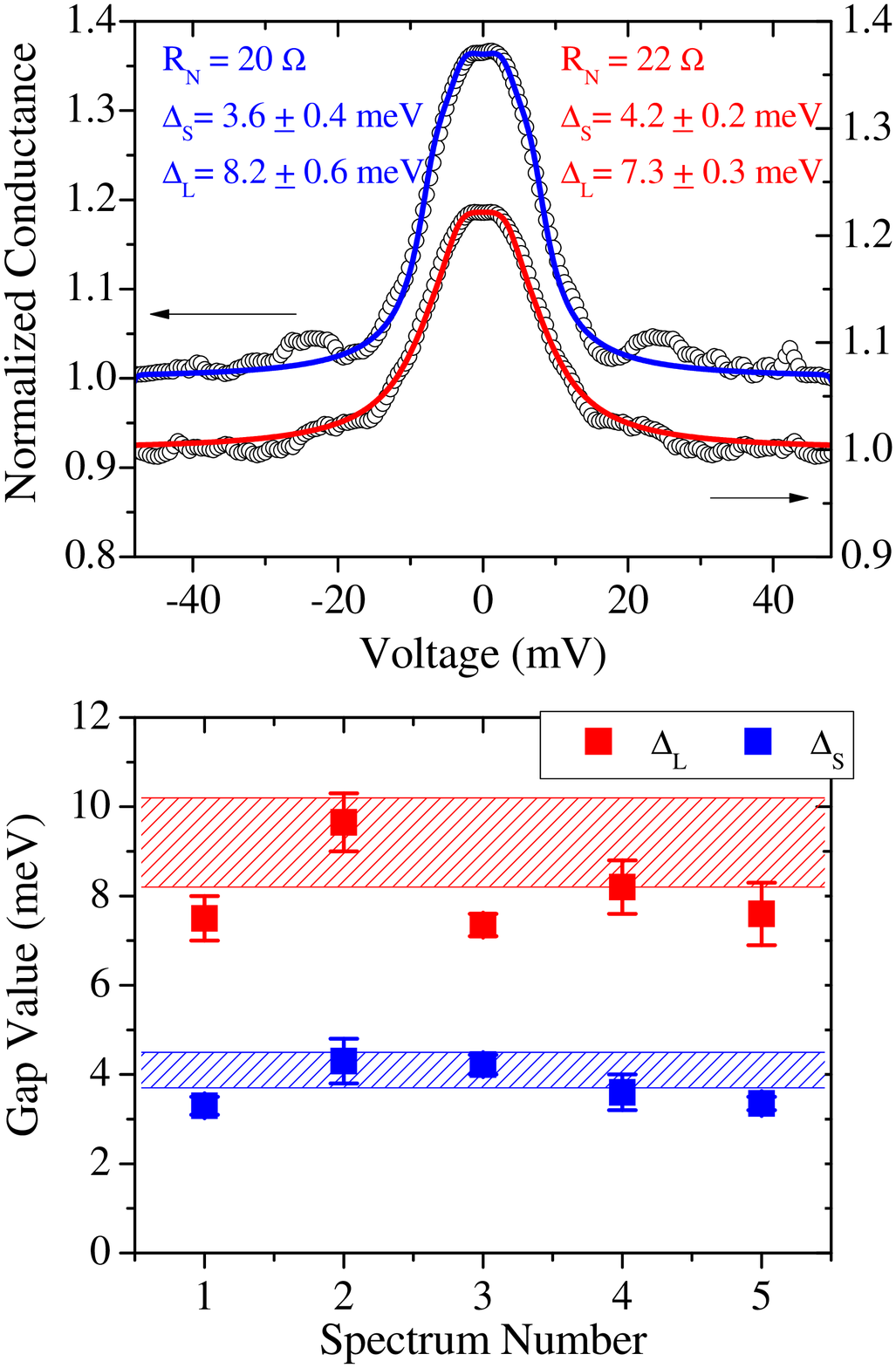}
\end{center}
\caption {}
\label{fig:PCARS}
\end{figure}

\bigskip
\textbf{Figure 3} \\ 
Examples of AFM measurements on the surface of a Co-doped Ba-122 thin film. (a) AFM image over
an area of $3.75 \,\mu\mathrm{m} \times 3.75\,\mu\mathrm{m}$, exhibiting a
clear pattern of flat, rectangular terraces of about 100 nm in size.
A comparison with the FESEM image of Figure \ref{fig:FESEM}a
(taken on a similar scale) shows that the morphological
information provided by FESEM and AFM is perfectly consistent; note that the
latter does not show the stripe-like pattern seen in Figure \ref{fig:FESEM}c.
(b) AFM image over a smaller area of $2 \,\mu\mathrm{m} \times 2\,\mu\mathrm{m}$
and thus with a greater magnification. (c) The same image as in (b) but after application of a Prewitt horizontal gradient filter so as to highlight the elevation of the terraces above the
background of well-connected grains.

\bigskip
\textbf{Figure 4} \\ 
Summary of structural characterization by X-ray diffraction.
(a) $\theta-2\theta$ scans in the Bragg-Brentano geometry with Co-K$\alpha$
radiation; (b) $\phi$-scans in a texture goniometer with Cu-K$\alpha$
radiation for the 112 reflection of Co-doped Ba-122 and the 111 reflection
of CaF$_2$.

\bigskip
\textbf{Figure 5} \\ 
Transport and magnetic measurements on a Co-doped Ba-122 thin film.
(a) The local critical temperature determined by PCARS (hollow symbols) compared to the resistive transition of the same film ({\color{blue} red filled} symbols). Left inset: a drawing of the film indicating the contacts used for the resistance measurement ({\color{blue} red} dots) and the regions where the point contacts were made ({\color{blue}white} dots). Right inset: the normalized resistance $R(T)/R(250\mathrm{K})$ compared to that of slightly underdoped single crystals
(from \cite{chu09}). (b) Temperature dependence of the normalized magnetic moment $m(T)/m(1.8\mathrm{K})$  of the film ({\color{blue} blue} open symbols) compared to the resistive transition of the same film ({\color{blue} red filled} symbols).

\bigskip
\textbf{Figure 6} \\ 
Summary of critical current measurements on a Co-doped Ba-122 thin film.
(a) Field dependence (up to 9 T) of the critical current density
$J\ped{c}$ for $H \parallel c$ (open symbols) and $H \parallel ab$ (solid symbols)
at various temperatures. (b) Angular dependence of $J\ped{c}$ measured
at $T=14$ K and in the presence of magnetic fields of different intensity, from 1 T to 9 T.

\bigskip
\textbf{Figure 7} \\ 
Examples of PCARS results on a Co-doped Ba-122 thin film.
(a) Two examples of normalized PCARS spectra measured in different points
of the same $8\%$ Co-doped film (symbols) with the relevant 2D BTK fit (lines). The values of the {\color{blue} gap amplitudes $\Delta\ped{L}$ and $\Delta\ped{S}$} are indicated in the labels.
(b) Summary of gap values obtained in different spectra in the same film (squares). The horizontal dashed bands represent the range of gap values found in single crystals with the same doping content.

\newpage

\newpage

\newpage

\newpage

\newpage

\newpage

\newpage

\begin{thebibliography}{10}
\expandafter\ifx\csname url\endcsname\relax
  \def\url#1{\texttt{#1}}\fi
\expandafter\ifx\csname urlprefix\endcsname\relax\def\urlprefix{URL }\fi
\expandafter\ifx\csname href\endcsname\relax
  \def\href#1#2{#2} \def\path#1{#1}\fi

\bibitem{johnston10}
D.~Jonhston, The puzzle of high temperature superconductivity in layered iron
  pnictides and chalcogenides, Advances in Physics 59 (2010) 803.

\bibitem{kadin99}
A.~L. Kadin, {Introduction to Superconducting Circuits}, Wiley-Interscience,
  1999.

\bibitem{seidel11}
P.~Seidel, Josephson effects in iron based superconductors, Supercond. Sci.
  Technol. 24 (2011) 043001.

\bibitem{katase09}
T.~Katase, H.~Hiramatsu, H.~Yanagi, T.~Kamiya, M.~Hirano, H.~Hosono,
  {Atomically-flat, chemically-stable, superconducting epitaxial thin film of
  iron-based superconductor, cobalt-doped $\mathrm{BaFe_2As_2}$}, Solid State
  Comunications 149 (2009) 2121--2124.

\bibitem{iida09}
K.~Iida, J.~H\"{a}nisch, R.~H\"{u}hne, F.~Kurth, M.~Kidszun, S.~Haindl,
  J.~Werner, L.~Schultz, B.~Holzapfel, {Strong T$_{C}$ dependence for strained
  epitaxial $\mathrm{Ba(Fe_{1-x}Co_x)_{2}As_{2}}$ thin films}, Appl. Phys. Lett
  95 (2009) 192501.

\bibitem{lee10b}
S.~Lee, J.~Jiang, Y.~Zhang, C.~Bark, C.~Weiss, J. D.and~Tarantini, C.~T.
  Nelson, H.~Jang, C.~M. Folkman, S.~H. Baek, A.~Polyanskii, D.~Abraimov,
  A.~Yamamoto, J.~Park, X.~Q. Pan, E.~E. Hellstrom, D.~C. Larbalestier, C.~B.
  Eom, {Template engineering of Co-doped $\mathrm{BaFe_{2}As_{2}}$
  single-crystal thin films}, Nature Mater. 9 (2010) 397.

\bibitem{engelmann13}
J.~Engelmann, V.~Grinenko, P.~Chekhonin, W.~Skrotzki, D.~Efremov, S.~Oswald,
  K.~Iida, R.~H\"uhne, J.~H\"anisch, M.~Hoffmann, F.~Kurth, L.~Schultz,
  B.~Holzapfel, {Strain induced superconductivity in the parent compound
  $\mathrm{BaFe_2As_2}$}, Nature Communications 4 (2013) 2877.

\bibitem{kurth13b}
F.~Kurth, E.~Reich, J.~H\"{a}nisch, A.~Ichinose, I.~Tsukada, R.~H\"{u}hne,
  S.~Trommler, J.~Engelmann, L.~Schultz, B.~Holzapfel, K.~Iida, {Versatile
  fluoride substrates for Fe-based superconducting thin films}, Appl. Phys.
  Lett. 102 (2013) 142601.

\bibitem{kurth13a}
F.~Kurth, K.~Iida, S.~Trommler, J.~H\"{a}nisch, K.~Nenkov, J.~Engelmann,
  S.~Oswald, J.~Werner, L.~Schultz, B.~Holzapfel, S.~Haindl, {Electronic phase
  diagram of disordered Co doped $\mathrm{BaFe_2As_{2}}$}, Supercond. Sci.
  Technol. 26 (2013) 025014.

\bibitem{koon96}
D.~Koon, C.~J. Knickerbocker, {Effects of macroscopic inhomogeneities on
  resistive and Hall measurements on crosses, cloverleafs, and bars}, Rev. Sci.
  Instrum. 67 (1996) 4282.

\bibitem{koon98}
D.~Koon, W.~K. Chan, {Direct measurement of the resistivity weighting
  function}, Rev. Sci. Instrum. 69 (1998) 4218.

\bibitem{vaglio93}
R.~Vaglio, C.~Attanasio, L.~Maritato, A.~Ruosi, {Explanation of the
  resistance-peak anomaly in nonhomogeneous superconductors}, Phys. Rev. B 47
  (1993) 15302.

\bibitem{mosqueira94}
J.~Mosqueira, A.~Pomar, A.~D\'{i}az, J.~A. Veira, F.~Vidal, {Resistivity
  anomalies above the superconducting transition in
  Y$_1$Ba$_2$Cu$_3$O$_{7-\delta}$ crystals and non-uniformly distributed
  critical-temperature inhomogeneities}, Physica C 34 (1994) 225.

\bibitem{chu09}
J.-H. Chu, J.~G. Analytis, C.~Kucharczyk, I.~R. Fisher, {Determination of the
  phase diagram of the electron-doped superconductor
  $\mathrm{Ba(Fe_{1-x}Co_x)_2As_2}$}, Phys. Rev. B 79 (2009) 014506.

\bibitem{pecchio13}
P.~Pecchio, D.~Daghero, G.~A. Ummarino, R.~S. Gonnelli, F.~Kurth, B.~Holzapfel,
  K.~Iida, {Doping and critical-temperature dependence of the energy gaps in
  Ba(Fe$_{1-x}$Co$_x$)$_2$As$_2$ thin films}, Phys. Rev. B 88 (2013) 174506.

\bibitem{daghero10}
D.~Daghero, R.~Gonnelli, {Probing multiband superconductivity by point-contact
  spectroscopy}, Supercond. Sci. Technol. 23 (2010) 043001.

\bibitem{andreev64}
A.~Andreev, {Thermal conductivity of the intermediate state of
  superconductors}, Zh. Eksp. Teor. Fiz. 46 (1964) 1823, engl. Transl. Sov.
  Phys.-JETP \textbf{19}, 1228 (1974).

\bibitem{iida10b}
K.~Iida, T.~H\"{a}nisch, T.~Thersleff, F.~Kurth, M.~Kidszun, S.~Haindl,
  R.~H\"{u}hne, L.~Schultz, B.~Holzapfel, Scaling behavior of the critical
  current in clean epitaxial $\mathrm{Ba(Fe_{1-x}Co_x)_2As_2}$ thin films,
  Phys. Rev. B 81 (2010) 100507(R).

\bibitem{putti10}
M.~Putti, I.~Pallecchi, E.~Bellingeri, M.~R. Cimberle, M.~Tropeano,
  C.~Ferdeghini, A.~Palenzona, C.~Tarantini, A.~Yamamoto, J.~Jiang,
  J.~Jaroszynski, F.~Kametani, D.~Abraimov, A.~Polyanskii, J.~D. Weiss, E.~E.
  Hellstrom, A.~Gurevich, D.~C. Larbalestier, R.~Jin, B.~C. Sales, A.~S. Sefat,
  M.~A. McGuire, D.~Mandrus, P.~Cheng, Y.~Jia, H.~H. Wen, S.~Lee, C.~B. Eom,
  {New Fe-based superconductors: properties relevant for applications},
  Supercond. Sci. Technol. 23 (2010) 034003.

\bibitem{chen10b}
T.~Y. Chen, S.~X. Huang, C.~L. Chien, {Pronounced effects of additional
  resistance in Andreev reflection spectroscopy}, Phys. Rev. B 81 (2010)
  214444.

\bibitem{doring13}
S.~D\"oring, S.~Schmidt, S.~Gottwals, , S.~Schmidl, V.~Tympel, I.~M\"onch,
  F.~Kurth, K.~Iida, B.~Holzapfel, P.~Seidel, {Influence of the spreading
  resistance on the conductance spectrum of planar hybrid thin film SNS'
  junctions based on iron pnictides}, unpublished, arXiv:1309.1641 (Sep 2013).

\bibitem{BTK}
G.~E. Blonder, M.~Tinkham, T.~M. Klapwijk, {Transition from metallic to
  tunneling regimes in superconducting microconstrictions: Excess current,
  charge imbalance, and supercurrent}, Phys. Rev. B 25 (1982) 4515.

\bibitem{kashiwaya00}
S.~Kashiwaya, Y.~Tanaka, {Tunnelling effects on surface bound states in
  unconventional superconductors}, Rep. Prog. Phys. 63 (2000) 1641–1724.

\bibitem{terashima09}
K.~Terashima, Y.~Sekiba, J.~H. Bowen, K.~Nakayama, T.~Kawahara, T.~Sato,
  P.~Richard, Y.-M. Xu, L.~J. Li, G.~H. Cao, Z.-A. Xu, H.~Ding, T.~Takahashi,
  {Fermi surface nesting induced strong pairing in iron-based superconductors},
  Proc. Natl. Acad. Sci (USA) 106 (2009) 7330.

\bibitem{tortello10}
M.~Tortello, D.~Daghero, G.~A. Ummarino, V.~A. Stepanov, J.~Jiang, J.~D. Weiss,
  E.~E. Hellstrom, R.~S. Gonnelli, {Multigap Superconductivity and Strong
  Electron-Boson Coupling in Fe-Based Superconductors: A Point-Contact
  Andreev-Reflection Study of Ba(Fe$_{1-x}$Co$_x$)$_2$As$_2$ Single Crystals},
  Phys. Rev. Lett. 105 (2010) 237002.

\end{thebibliography}
\end{document}